\documentclass[prl,twocolumn]{revtex4}
\usepackage{amsmath,amssymb}
\usepackage{amsthm}
\usepackage{epsfig}
\usepackage{xcolor}
\usepackage{fixltx2e}
\usepackage[normalem]{ulem}
\usepackage{siunitx}

\usepackage{xr}
\graphicspath{{eps/}}            

\begin{document}

\title{Hybridization of Kerr Solitons in Coupled Microresonators}

\author{Alena Kolesnikova$^{1}$}
\email[Corresponding author : ]{a.kolesnikova@skoltech.ru}
\author{Ivan Pshenichnyuk$^{1}$}
\author{Andrey Gelash$^{1}$}
\email[Corresponding author : ]{a.gelash@skoltech.ru}

\affiliation{$^{1}$Center for Engineering Physics, Skolkovo Institute of Science and Technology, Moscow, Russia}

\begin{abstract}
Recent advances in manufacturing photonic integrated devices enable efficient coupling between high-Q microresonators in both linear and nonlinear regimes, creating a tunable, complex, hybridized optical system. Considering two coupled microresonators with normal and anomalous dispersion and equal free spectral range (FSR), we theoretically predict a novel nonlinear phenomenon: fully coherent hybridization of dissipative Kerr solitons (DKS) and propose a realistic integrated photonic design for its experimental observation. Using the Lugiato-Lefever equations in the supermode basis, we show that the emergent picture of inter-resonator DKS interactions can be understood as the formation of coherent structures in both supermodes generated by an unusual four-wave mixing process. The found hybridized DKS states can exhibit a broad, flat spectral profile near the pumped mode and remarkable oscillatory features in the spectral wings, promising broad applications in the generation and control of optical Kerr frequency combs.
\end{abstract}
\maketitle
Hybridization of physical systems is a fundamental phenomenon responsible for the formation of complex structures in nature, as illustrated by the well-known omnipresent linear mixing of atomic orbitals \cite{ruedenberg1962physical}. Nonlinear interactions result in fundamentally different hybridized states emerging beyond the conventional superposition of the system components. In optics, the coupling of nonlinear soliton coherent states of light has recently been experimentally demonstrated, promising various applications in the development of photonic devices \cite{tikan2021emergent,yuan2023soliton,ji2024multimodality,letsou2025hybridized}. Recent progress in integrated photonics unlocked the development of chip-scale efficiently coupled high-Q optical microresonators with Kerr nonlinearity – from photonic dimers and trimers to large-scale mirroring lattices, which are becoming a physical testbed for discovering new hybridized states of light \cite{komagata2021dissipative,tikan2022protected,flower2024observation,xu2025chip}. A joint action of nonlinearity, dispersion, pump, and dissipation allows the generation of dissipative Kerr solitons (DKS) \cite{ankiewicz2008dissipative,Gorodetsky2018dissipative,rosanov2026dissipative}. Being first theoretically predicted using the Lugiato-Lefever equation (LLE) model \cite{lugiato1987spatial,kaup1978solitons,nozaki1986low,wabnitz1996control}, DKS have been the subject of intensive study since their experimental observation on optical resonator and microresonator platforms \cite{herr2014temporal,leo2010temporal, Leo2013Dynamics,lucas2017breathing}. Thanks to their exceptional stability, DKS are a reliable source of optical frequency combs. On-chip microresonators are capable of generating optical frequency microcombs in the GHz-THz range of repetition rates, proving versatile photonic integrated solutions in frequency metrology, telecommunications, and spectroscopy, to name a few \cite{Gorodetsky2018dissipative,herr2016dissipative}.

Although conventional single-ring DKS systems have been under intensive study over the past decade \cite{herr2026frequency}, unresolved issues regarding self-referencing, detectability and conversion efficiency of microcombs demand further advanced exploration of novel coherent dissipative soliton states with tailored optical frequency comb spectra profile \cite{anderson2022zero,lucas2023tailoring,wu2025ultraflat}. Recent advances in manufacturing photonic integrated coupled high-Q microresonators have enabled the generation of novel types of DKS and other light patterns, opening avenues for engineering the spectral properties of microcombs \cite{helgason2021dissipative,tikan2021emergent,tikan2022protected,yuan2023soliton,flower2024observation,liu2025near}. Here, we study a system of coupled microresonators A and B, schematically depicted in Fig.~1(a), where we indicate the pump photon flux $|s_{\text{in}}|^2$ and the resonators coupling coefficient $J$. By design, the microresonators have equal free spectral range (FSR) and opposite signs of the quadratic mode dispersion $D_2$, that is controlled by the microring waveguide geometry. The DKS optical spectra can be measured from both outports of A and B.
\begin{figure}[t!]
    \centering
    \includegraphics[width=1.0\linewidth]{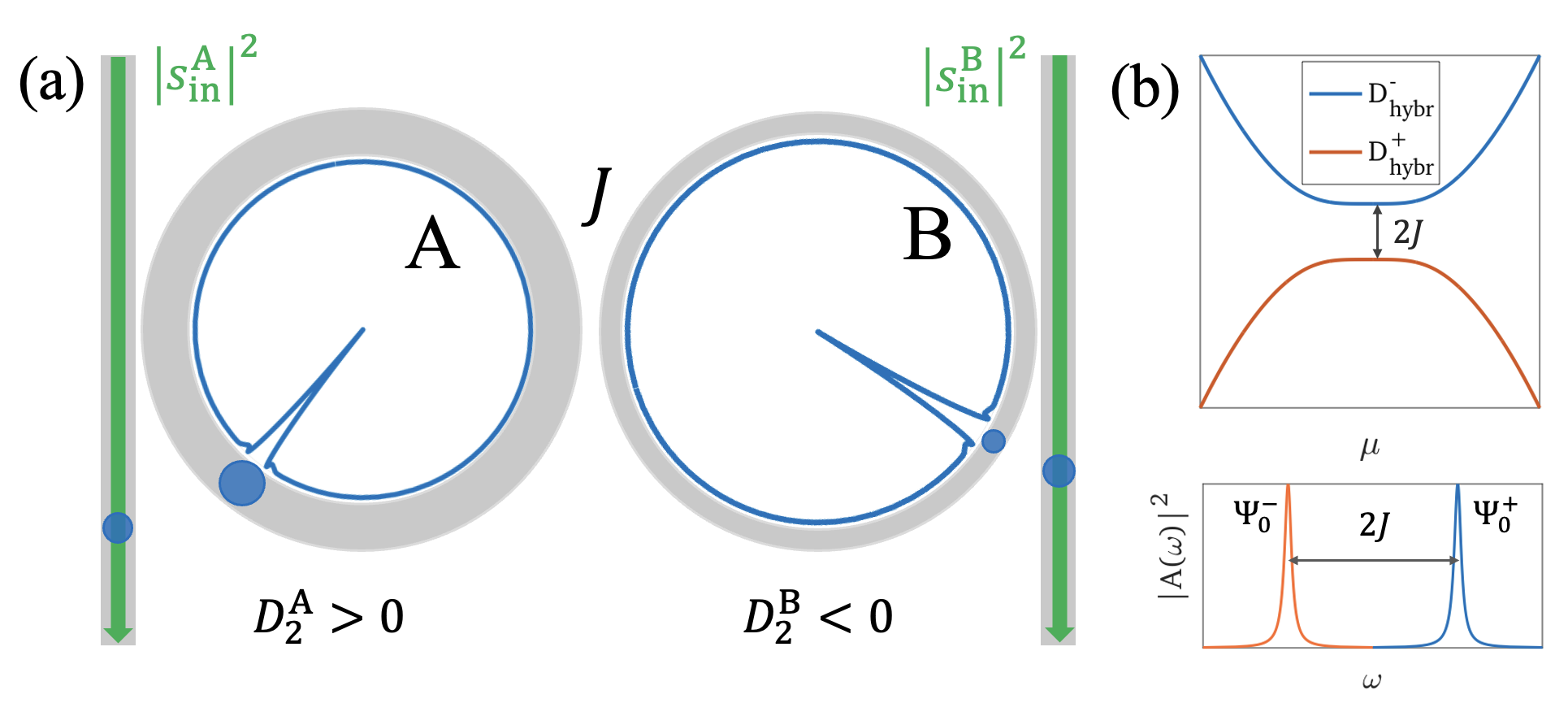}
    \caption{(a) Schematic representation of the coupled microresonators with DKS in each ring produced by continuous wave pump and inter-resonator interaction. (b) An example of its hybridized dispersion and intracavity power (split resonances) in linear regime at \(\Delta_{\text{res}}=0\).}
    \label{fig:1}
\end{figure}

Nonlinear dynamics in the resonator is determined by the sign of $D_2$. In the anomalous dispersion regime ($D_2>0$), modulation instability leads to self-organization of Turing roll and soliton patterns \cite{qi2019dissipative}, while in the regime of normal dispersion ($D_2<0$), an external trigger is required to form dark soliton pulses \cite{xu2021frequency,anderson2022zero}. These coherent nonlinear structures have been widely observed experimentally on different optical platforms \cite{grelu2015nonlinear,Gorodetsky2018dissipative,herr2026frequency} and investigated theoretically using direct numerical simulations of the LLE and analysis of its solution branches bifurcations \cite{Parra-Rivas2018localized,parra2018bifurcation}. Recently, a nonlinear regime of operation has been achieved experimentally for anomalous-anomalous, normal-normal, and normal-anomalous coupled microresonator systems \cite{tikan2021emergent,yuan2023soliton,pidgayko2023voltage}. Mutual linear influence of the resonators splits the dispersions of the same signs, and the resulting coupled system can host such coherent states as gear solitons and soliton hopping, which can be efficiently analyzed in the supermode basis of the coupled LLE model \cite{tikan2021emergent,komagata2021dissipative}. Dispersion of anomalous-normal coupled resonators hybridizes non-trivially, forming complex, tunable landscapes \cite{pidgayko2023voltage}; however, DKS in such systems are still missing both in experiment and theory.

Hereby, through direct numerical simulations of the coupled LLE model, we discover coherent DKS living in hybridized landscapes of anomalous and normal dispersions, nonlinearly hybridized by the Kerr effect. The latter means that in the supermode basis of the system, both symmetric and antisymmetric components contain coherent structures, and the resulting soliton states cannot be described by an effective single LLE model. In contrast to hybridization of solitons observed in anomalous-anomalous microresonators \cite{tikan2021emergent,komagata2021dissipative}, here both supermodes can support coherent pulses, leading to an unusual shape and spectra of the output resonator fields. Practically important, the linear hybridization of anomalous and normal dispersions forms a broad, flat region around the pump mode, resulting in a flat-topped spectrum of the hybridized solitons, which is widely tunable via the interresonator detuning parameter.

In linear regime, the coupled resonators A and B are described using coupled equations for azimuthal modes,
\begin{multline}
\label{Eq:coupled-mode_A-B}
    \cfrac{\partial A_{\mu}^{i}}{\partial t}=\left[-\cfrac{\kappa^{i}}{2}-i\Delta^{i}-i\cfrac{D_2^{i}}{2}\mu^2\right]A_\mu^{i} +iJA_\mu^{j}\\+\delta_{0\mu}\sqrt{\kappa^{i}_{c}}s^{i}_{\text{in}}, \quad\quad i,j=\{\text{A,B}\},\quad i\neq j.
\end{multline}
where in addition to parameters shown in Fig.~\ref{fig:1} we introduce the intracavity losses \(\kappa^{i}\), the coupling coefficient between resonator and input/output waveguide \(\kappa_c^{i}\) and the pump detunings from microresonators resonant zero-mode frequency: \(\Delta^{\text{A}}=\omega^{\text{A}}_0-\omega_p=\Delta_0+\Delta_{\text{res}}/2\), \(\Delta^{\text{B}}=\omega^{\text{B}}_0-\omega_p=\Delta_0-\Delta_{\text{res}}/2\). The pump vector \(\boldsymbol{s}_{\text{in}}=\left(s_{\text{in}}^{\text{A}},s_{\text{in}}^{\text{B}}\right)\) defines power and phase of the input continuous wave laser field and is related to the total pump power $P_{\text{in}}$ as $|s^{\text{A}}_{\text{in}}|^2 + |s^{\text{B}}_{\text{in}}|^2 = P_{\text{in}}/\hbar\omega_p$.

The resonator's coupling results in the splitting of the azimuthal modes \(\mu\) into a set of supermodes \(\Psi^{\pm}_\mu\) characterized by hybridized dispersion coefficients \(D_{\text{hybr}}^{\pm}(\mu)\) and hybridized spectrally depended losses \(\kappa_{\text{hybr}}^\pm(\mu)\), following from diagonalization of system (\ref{Eq:coupled-mode_A-B}); see the Supplementary materials~\cite{SuppM} for details and the full expressions. We preform our main analysis for the case \(\kappa^{\text{A}}=\kappa^{\text{B}}\), when the hybridized losses become spectrally independent \(\kappa_{\text{hybr}}^\pm\equiv\c (\kappa^{\text{A}}+\kappa^{\text{B}})/2\), and the coupled mode equations (\ref{Eq:coupled-mode_A-B}) for supermodes can be written as:
\begin{equation}
\label{Eq:coupled-mode_AS-S}
    \cfrac{\partial \Psi^{\pm}_\mu}{\partial t}=\left[-\cfrac{\kappa_{\text{hybr}}^\pm}{2}-i\Delta_{0}-i\cfrac{D_{\text{hybr}}^{\pm}(\mu)}{2}\right]\Psi^{\pm}_\mu+\delta_{0\mu}F^\pm_{\text{in}},
\end{equation}
where $F^\pm_{\text{in}}$ are the supermode pump vector $\boldsymbol{F}_{\text{in}}$ components and the hybridized dispersion
\begin{multline}
\label{Eq:Dhybr}
   D_{\text{hybr}}^\pm(\mu) = \cfrac{D_2^{\mathrm{A}}+D_2^{\mathrm{B}}}{2}\mu^2\mp\\\sqrt{\left(\Delta_{\text{res}}+\cfrac{D_2^{\text{A}}-D_2^{\text{B}}}{2}\mu^2\right)^2+4J^2}.
\end{multline}
Note that resonator losses can be adjusted by changing the position of the bus waveguides. In the supplementary materials~\cite{SuppM}, we demonstrate that setting different loss values does not lead to a significant change in the dispersive hybridization and nonlinear dynamics of the system, and the results obtained for the case \(\kappa^{\text{A}}=\kappa^{\text{B}}\) underline the general behavior of the system.

\begin{figure}[t!]
    \centering
 \includegraphics[width=\linewidth]{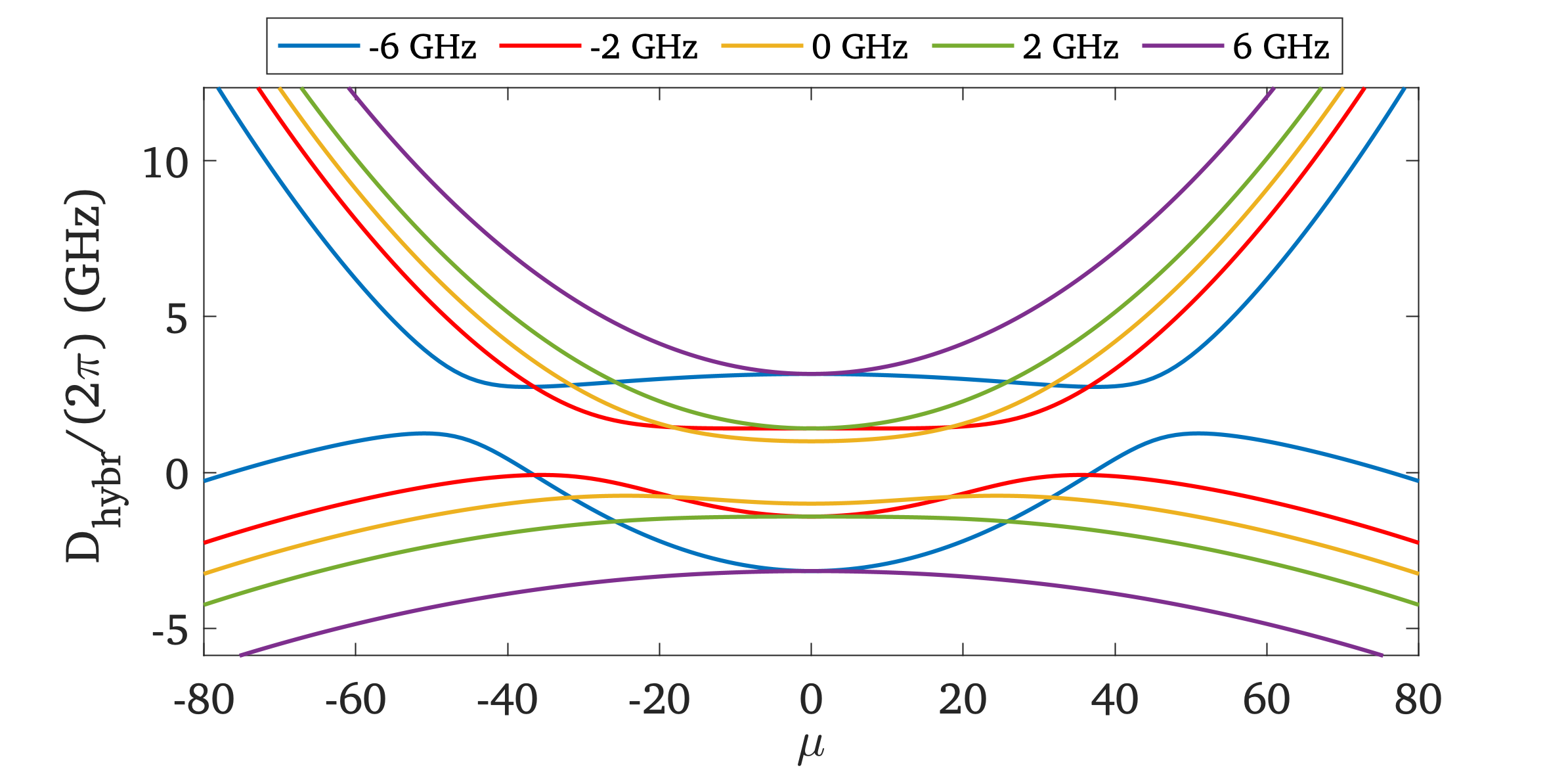}
    \caption{Hybridized dispersion (\ref{Eq:Dhybr}) at \(D_2^{\text{A}}/(2\pi) = 5\) MHz, \(D_2^{\text{B}}/(2\pi) = -1\) MHz, \(J/(2\pi)=1\,\) GHz and different \(\Delta_{\text{res}}\).}
    \label{fig:2}
\end{figure}

The linear transformation connecting the bases of resonators $A_\mu^{\text{A,B}}$ and supermodes $\Psi^{\pm}_\mu$ can be found as,
\begin{eqnarray}
\label{Eq:direct-tranformation}
\Psi^{\pm}_\mu &=& \cfrac{A_\mu^{\text{A}}+C^{\pm}_{\mu} A_\mu^{\text{B}}}{\sqrt{1+|C^{\pm}_{\mu}|^2}}\,,
\\
\label{Eq:inverse-tranformation-A}
A^{\text{A}}_\mu &=& \cfrac{C^-_{\mu}\sqrt{1+|C^+_{\mu}|^2}\Psi^{+}_\mu-C^+_{\mu} \sqrt{1+|C^-_{\mu}|^2}\Psi^{-}_\mu}{C^-_{\mu}-C^+_{\mu}},
\\
\label{Eq:inverse-tranformation-B}
A^{\text{B}}_\mu &=& \cfrac{-\sqrt{1+|C^+_{\mu}|^2}\Psi^{+}_\mu+ \sqrt{1+|C^-_{\mu}|^2}\Psi^{-}_\mu}{C^-_{\mu}-C^+_{\mu}},
\end{eqnarray}
where
\begin{multline}
    C^{\pm}_{\mu} = \left(\cfrac{\Delta_{\text{res}}}{2}+\cfrac{D_2^{\text{A}}-D_2^{\text{B}}}{4}\mu^2\pm\right.\\\left.\cfrac{1}{2}\sqrt{\left(\Delta_{\text{res}}+\cfrac{D_2^{\text{A}}-D_2^{\text{B}}}{2}\mu^2\right)^2+4J^2}\right)/J.
\label{Eq:A2}
\end{multline}
see the Supplementary materials~\cite{SuppM} for details.

    \begin{figure*}[ht!]
        \centering
        \includegraphics[width=1\linewidth]{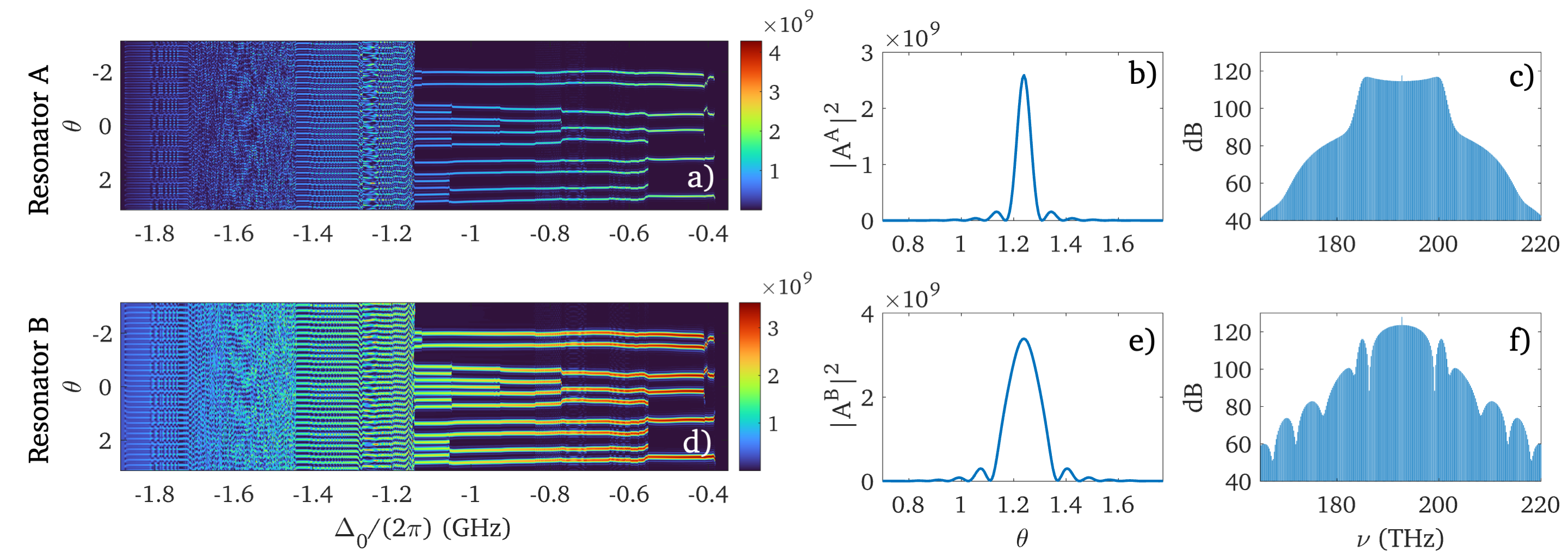}
        \caption{Numerical modeling of hybridized soliton states generation in two coupled microresonators with anomalous (Resonator A) and normal (Resonator B) dispersion. a,d) Evolution of the spatial distribution of the field \(|A^{\text{A,B}}(\theta,\Delta_0(t))|^2\) inside microresonators A and B obtained in numerical simulations of the model (\ref{Eq:coupled-LLE}) with parameters: $\text{FSR} = 197$ GHz, \(D_2^{\text{A}}/(2\pi) = 5\) MHz, \(D_2^{\text{B}}/(2\pi) = -1\) MHz, \(J/(2\pi) = 1\) GHz, \(\kappa/(2\pi)=2\kappa_c/(2\pi) = 100\) MHz, \(\Delta_{\text{res}}/(2\pi)=-3.2\,\)GHz, \(P_{\text{in}}=0.2\,\)
        W. (b, e) Fully coherent hybridized soliton state at \(\Delta_0/(2\pi)=-0.5\, \)GHz and (c, f) its corresponding optical frequency comb in each microresonator.}
        \label{fig:3}
    \end{figure*}

    \begin{figure*}[ht!]
    \centering
    \includegraphics[width=1\linewidth]{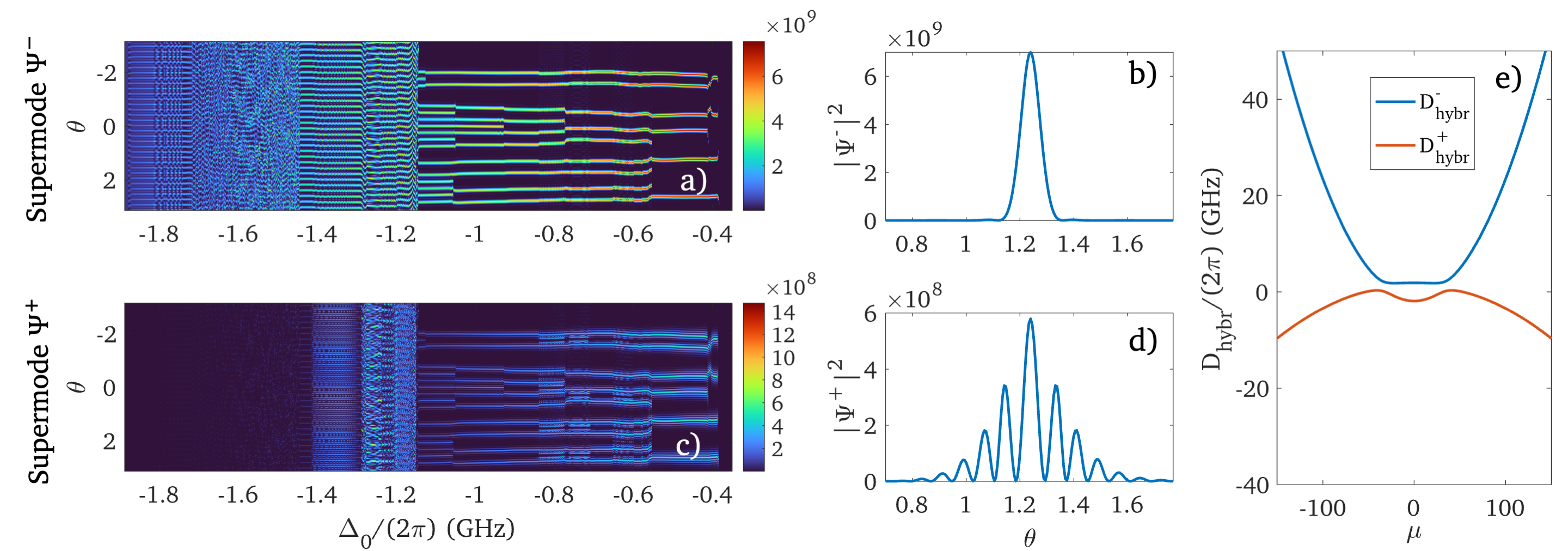}
    \caption{The same numerical modeling as shown in Fig.~\ref{fig:3} presented in the basis of supemodes. a,c) Evolution of the spatial distribution of the field \(|\Psi^{\pm}(\theta,\Delta_0(t))|^2\). 
    b, d) Fully coherent supermodes of the hybridized soliton state at \(\Delta_0/(2\pi)=-0.5 \,\)GHz. e) Hybridized dispersion of the supermodes according to Eq.~(\ref{Eq:Dhybr}).}
    \label{fig:4}
    \end{figure*}

The hybridized dispersion (\ref{Eq:Dhybr}) is determined by geometry of the coupled system (Fig.~\ref{fig:1}(a)) and, in addition, can be widely tuned using relative shift of the resonant frequencies \(\Delta_{\text{res}}\), which can be controlled by heating the microresonators \cite{tikan2021emergent,pidgayko2023voltage}. When microrings are identical \(D_2^{\text{A}}=D_2^{\text{B}}\), each pair of modes is split with a shift \(2J\) resulting in two identical parabolas and the dispersion hybridization is absent, see Eq.~(\ref{Eq:Dhybr}). In contrast, when \(D_2^{\text{A}}=-D_2^{\text{B}}\) and \(\Delta_{\text{res}}=0\), the hybridization is manifested as \(D_{\text{hybr}}^\pm=\mp\sqrt{\left(D_2^{\text{A}}\mu^2\right)^2+J^2}\), which cannot be realized when the resonators are separated. In this case, quadratic contribution into the hybridized dispersion vanishes for \(\mu \ll J/D_2^{\text{A}}\), forming a broad flat area near the pump mode, so that the set of symmetric supermodes \(\Psi^+_\mu\) has an effective negative mode dispersion, while the set of antisymmetric supermodes \(\Psi^-_\mu\) has a positive mode dispersion; see an illustration in Fig.~\ref{fig:1}(b). Remarkably, taking \(|D_2^{\text{A}}|\ne |D_2^{\text{B}}|\) in combination with variations of \(\Delta_{\text{res}}\) enables nontrivial dispersion hybridization landscapes with several local minima, see Fig.~\ref{fig:2}.

\begin{figure*}[t!]
    \centering
    \includegraphics[width=1\linewidth]{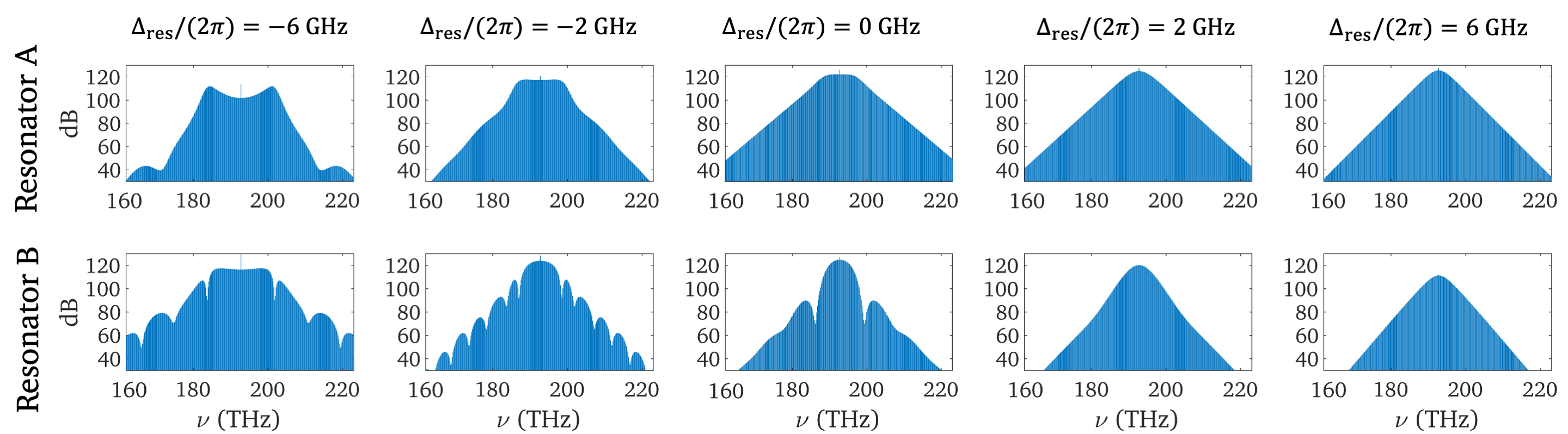}
    \caption{Optical frequency combs spectra of hybridized soliton states generated by scanning of the pump frequency at different relative shift of the resonant frequencies \(\Delta_{\text{res}}\) between the microresonators A and B. Other parameters are the same as used for numerical simulations shown in Fig.~\ref{fig:3}.}
   \label{fig:5}
\end{figure*}

As shown in \cite{pidgayko2023voltage}, a system of coupled microresonators with opposite dispersion signs can be experimentally implemented on a thick silicon nitride platform \cite{pfeiffer2018photonic} by making widths of the resonator waveguides different from each other. Based on numerical mode analysis of the microring waveguides and 3D finite element modeling of electromagnetic fields, we propose realistic designs of coupled microresonators with strongly hybridized dispersion and, as shown below, nonlinearly hybridized coherent soliton states; see details in the Supplementary materials~\cite{SuppM}. As an example, for waveguides A and B with dimensions $1400\times 800$ nm$^2$ and $922\times 800$ nm$^2$ and the curvature radii $R^\text{A}=112.5$ $\mu$m and $R^\text{B}=113.5$ $\mu$m, we obtain the dispersion coefficients at the pump freqency \(\omega/(2\pi)=193\) THz: \(D_2^{\text{A}}/(2\pi) = 5\) MHz, \(D_2^{\text{B}}/(2\pi) = -1\) MHz, so that microresonators A and B have anomalous and normal dispersions respectively. The slight difference between $R^\text{A}$ and $R^\text{B}$ provide equal FSR of the rings close to 197 GHz. We also find that in this configuration the coupling between
$\mathrm{TE}_0$ modes of the curved waveguides is very efficient and can be varied in the range up to hundred of GHz. For the following nonlinear numerical simulations we choose microresonator coupling strength \(J/(2\pi) = 1\) GHz. We also set \(\kappa/(2\pi)=2\kappa_c/(2\pi) = 100\) MHz and adjust the pump \(P_{\text{in}}=0.2\,\)W into an antisymmetric supermode near its resonant frequency detuning \(D_{\text{hybr}}^-(0)=\sqrt{\Delta_{\text{res}}^2/4+J^2}\), see Fig.~\ref{fig:1} where we indicate the split resonances. Note that pumping solely anomalous ring result in similar nonlinear dynamics, see details on the choice of the pump vector $\boldsymbol{s}_{\text{in}}$.

To study nonlinear dynamics of our coupled microresonators system we use the Lugiato-Lefever model written for the intracavity field distribution $A^{i}(\theta,t)=\sum_{\mu\in\mathbb{Z}} A^{i}_{\mu}(t)e^{-i\mu\theta}$ as follows:
\begin{multline}
\label{Eq:coupled-LLE}
    \cfrac{\partial A^{i}}{\partial t}=\left[-\cfrac{\kappa}{2}-i\Delta^{i}+iD_2^{i}\cfrac{\partial^2}{\partial\theta^2}+ig^{i}\left|A^{i}\right|^2\right]A^{i} \\+iJA^{j}+\sqrt{\kappa_c}s_{\text{in}}^{i},
\end{multline}
 where \(g^i\) are the nonlinear coefficients. Using standard technique to reach DKS states – scanning of the pump frequency with pump power above threshold of the modulation instability \cite{herr2014temporal}, we observed formation of Turing rolls and then chaotic field, which resulted in the emergence of novel coherent soliton states existing in both resonators in the red detuned region, see Fig.~\ref{fig:3} and Supplementary materials for more examples at different system parameters and pump power~\cite{SuppM}. Soliton A from the anomalous dispersion ring has a narrow profile with oscillating tails, which is due to flat region of $D_{\text{hybr}}^{-}$ close to quartic (fourth-order dispersion) solitons of single LLE model \cite{parra2022quartic,taheri2019quartic,blanco2016pure}. The optical frequency-comb spectrum of soliton A is nearly parabolic around the pump mode, with high-intensity lines; see Fig.~\ref{fig:3}(b,c). Soliton B from the normal dispersion ring is characterized by an unusually wide shape and oscillating frequency-comb spectrum (Fig.~\ref{fig:3}(e,f)). Remarkably, formation of these hybridized soliton sates cannot be described by a single effective LLE model, which follows from analysis performed in the basis of supermodes $\Psi^{\pm}(\theta,t)=\sum_{\mu\in\mathbb{Z}} \Psi^{\pm}_{\mu}(t)e^{-i\mu\theta}$, see Fig.~\ref{fig:4} and Supplementary materials \cite{SuppM} for other examples. Here $\Psi^{\pm}_{\mu}$ is obtained via linear transformation (\ref{Eq:direct-tranformation}) from the Fourier components of the LLE (\ref{Eq:coupled-LLE}) solution $A^{\text{A,B}}_{\mu}(t) = \int A^{\text{A,B}}(\theta,t)e^{i\mu\theta} d\theta$. 

Pumping energy into $\Psi^{-}$ supermode in the blue-detuned region of the resonance, we trigger the MI development and conventional formation of primary comb states described by an effective single LLE model with dispersion $D_{\text{hybr}}^{-}(\mu)$; see details in Supplementary materials~\cite{SuppM}. However, nonconventional four-wave mixing between supermodes, similar to that observed recently in coupled anomalous dispersion microresonators \cite{komagata2021dissipative}, unlocks energy transfer from $\Psi^{-}$ to $\Psi^{+}$ supermodes, which intensifies with detuning, see Fig.~\ref{fig:4}(c). When the impact of $\Psi^{+}$ reaches significant values, the effective single LLE model becomes invalid. The nonlinear interactions between supermodes result in the formation of two fully coherent dissipative solitons living in both supermodes (Fig.~\ref{fig:4}(b,d)). $\Psi^{-}$ soliton keeping main part of the energy and has a shape close to classical DKS, see (Fig.\ref{fig:4}(b)). Meanwhile, the narrow region of positive dispersion in $D_{\text{hybr}}^{+}(\mu)$, see Fig.~\ref{fig:4}(e), and coupling to $\Psi^{-}$ soliton generates an unusual soliton with oscillating profile in the supermode $\Psi^{+}$.

According to reverse transformations (\ref{Eq:inverse-tranformation-A}-\ref{Eq:inverse-tranformation-B}), the impact of supermodes $\Psi^{\pm}_{\mu}$ into the spectrum of resonators A and B (Fig.~\ref{fig:3}(c,d)) is determined by the value of $|C^{\mp}(\mu)|$ respectively. At considering parameters, direct evaluation of Eq.~(\ref{Eq:A2}) gives that $|C^{+}(\mu)|\lessgtr 1$ and $|C^{-}(\mu)|\gtrless 1$ at $\mu\lessgtr 30$, meaning that $\Psi^{-}$ supermode is responsible for the formation of the flat-top spectrum in the resonator A, while $\Psi^{-}$ supermode shapes oscillating wings in the spectrum of the resonator B.

\begin{figure}[t!]
    \centering
    \includegraphics[width=\linewidth]{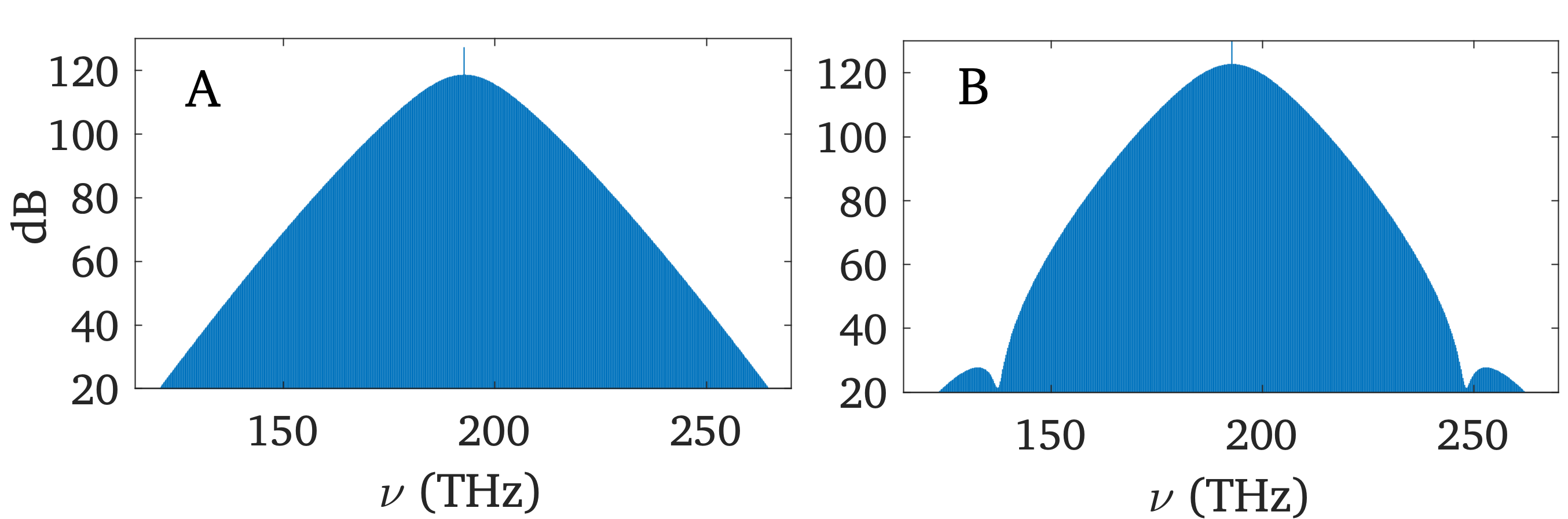}
    \caption{Octave-spanning spectra of hybridized solitons at strong coupling \(J/(2\pi)=100\,\)GHz and far-off \(\Delta_{\text{res}}=99\,\)GHz. 
    Other parameters are the same as in Fig.~\ref{fig:3}.}
    \label{fig:6}
\end{figure}

Relative shift of the resonant frequencies \(\Delta_{\text{res}}\) between the microresonators A and B is the pivotal parameter in control of optical frequency comb spectra of hybridized solitons, as we illustrate in Fig.~\ref{fig:5}. Changing \(\Delta_{\text{res}}\) towards zero and then positive values, we narrow the region of the flat-top spectrum in resonator A and reduce the number of oscillations in the frequency comb of resonator B. At \(\Delta_{\text{res}}/(2\pi)=2\) GHz, the flat spectrum region disappears, the hybridized dispersion acquires a parabolic shape (Fig.~\ref{fig:2}), and the soliton spectrum becomes similar to the classical spectrum in a single resonator (Fig.\ref{fig:5}). We observe the formation of hybridized solitons over a wide range of pump powers, beginning at 0.15 W, see Supplementary materials \cite{SuppM}. With an increase in power to 0.45 W, the region of hybridized solitons' existence is followed by bifurcation, leading to the formation of a dispersive wave in the symmetric supermode that results in a complex, full coexistence of solitons and frozen radiation, which provides an additional degree of freedom to tailor the hybridized frequency combs, see Fig.~\ref{fig:7} in the Appendix. Our numerical modeling of the coupling section between normal and anomalous microresonators (see the Supplementary materials~\cite{SuppM}) shows that $J$ can be of the order of FSR values when the distance between the waveguides is $\sim 50$ nm. In such a case, proper tuning of \(\Delta_{\text{res}}\) allows reaching optical frequency comb spectra of hybridized solitons fitting the full octave, see Fig.\ref{fig:6}.

In this work, we predicted the existence of a novel type of dissipative Kerr soliton in a system of two coupled microresonators with different waveguide widths, characterized by hybridized, widely tunable normal-anomalous dispersion. In contrast to the recent studies on gear solitons coexisting with dispersive waves and soliton hopping in systems of coupled resonators \cite{tikan2021emergent,komagata2021dissipative}, here we observe the formation of fully coherent structures in both supermodes, which sustain each other via the four-wave mixing process. Though we considered a specific example of the coupled microresonator system, previous experimental studies \cite{pidgayko2023voltage} and our numerical modeling suggest that a broad range of coupled microresonator geometries can support hybridized normal-anomalous dispersion and hybridized solitons.  A fundamental question for further studies is whether similar hybridized fully coherent Kerr soliton states can exist in the presence of Vernier effect \cite{yuan2023soliton}, in lattices of microresonators \cite{tusnin2023nonlinear,flower2024observation}, and in the regime of dark soliton pulses.

\begin{acknowledgments}
This work was supported by Russian Science Foundation project N\textsuperscript{\underline{\scriptsize o}} 25-72-31023.
\end{acknowledgments}


%

\clearpage
\onecolumngrid
\section{Appendix}

\begin{figure}[h!]
    \centering
    \includegraphics[width=1\linewidth]{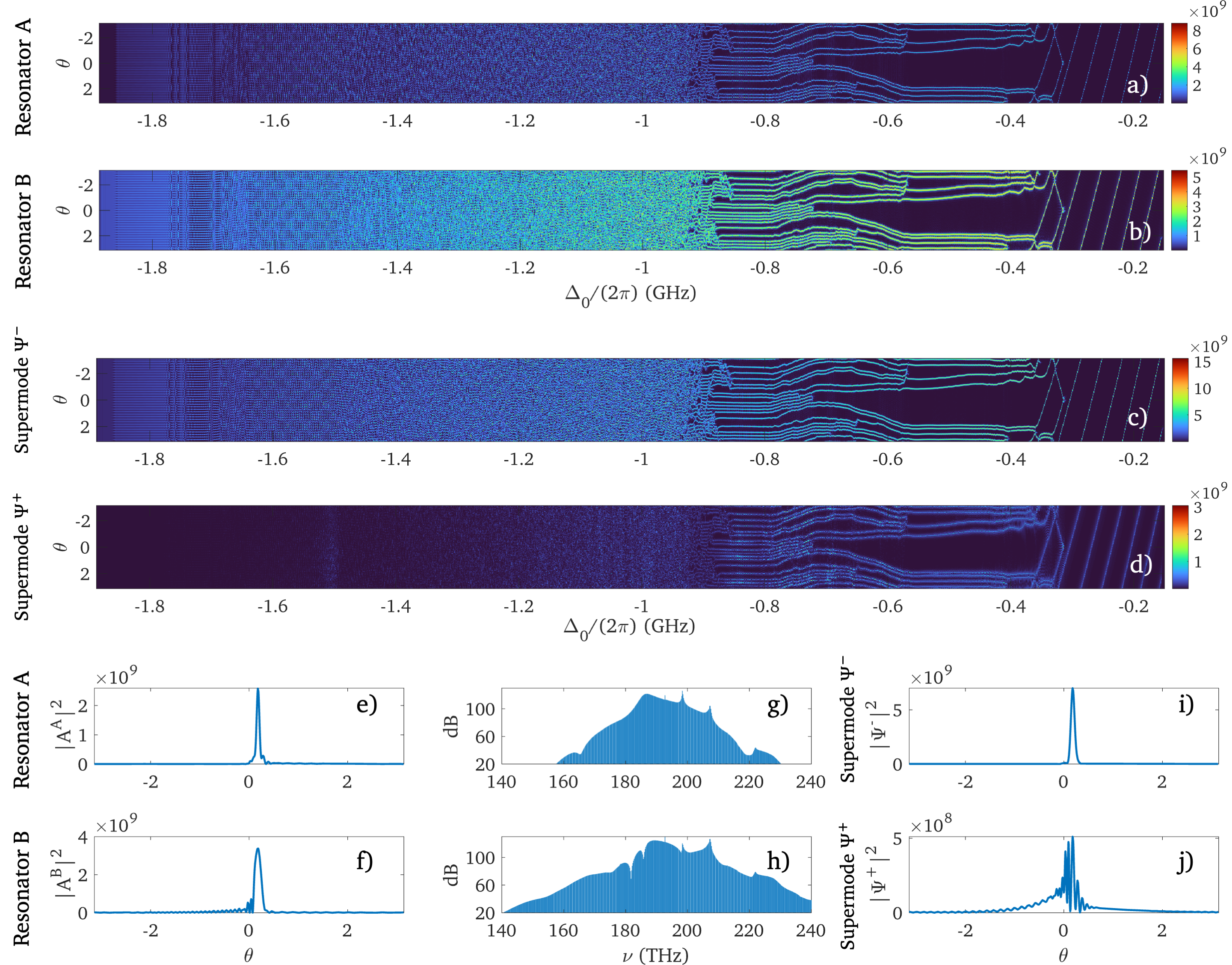}
    \caption{Numerical modeling of the coupled microresonator system from Figs.~\ref{fig:3}-\ref{fig:4} at higher pump power \(P_{\text{in}}=0.45\,\) W. Hybridized solitons undergo a bifurcation leading to formation of dispersive wave in symmetric supermode at \(\Delta_0/(2\pi) \approx -0.35\, \)GHz. a,b) Evolution of the spatial distribution of the the intensity \(\left|A^{\text{A,B}}(\theta,\Delta_0(t))\right|^2\) inside microresonators and c,d) in the supermode basis \(\left|\Psi^{\mp}(\theta,\Delta_0(t))\right|^2\). Fully coherent hybridized soliton states with dispersive wave at \(\Delta_0/(2\pi)=-0.25\, \)GHz in  e,f) the microresonator and i,j) supemode basis and g, h) corresponding optical frequency comb in each microresonator.}
    \label{fig:7}
\end{figure}

\end{document}